%% file: placement.tex
\documentclass[10pt, conference, letterpaper]{IEEEtran}
\IEEEoverridecommandlockouts
\usepackage{cite}
\usepackage{amsmath,amsthm,amssymb,amsfonts}
\usepackage[ruled,linesnumbered]{algorithm2e}
\DontPrintSemicolon
\usepackage{graphicx}
\usepackage{graphbox}
\usepackage{textcomp}
\usepackage{xcolor}
\usepackage{mathtools}
\usepackage{array}
\usepackage{float}
\usepackage{caption}
\usepackage{makecell}
\usepackage[labelformat=simple]{subcaption}
\usepackage{kantlipsum} 
\usepackage{mwe} 
\usepackage{xurl} 
\usepackage{hyperref}
\usepackage{booktabs}
\usepackage{tikz}
\usepackage{multicol}
\usepackage{multirow}

\usepackage{listings}
\usepackage{xcolor}
\definecolor{codegray}{RGB}{0,92,240}
\definecolor{codepurple}{rgb}{0.58,0,0.82}
\definecolor{keyword}{RGB}{186, 45, 162}
\lstdefinestyle{mystyle}{
    backgroundcolor=\color{white},   
    commentstyle=\color{codegray},
    keywordstyle=\bfseries\color{keyword},
    numberstyle=\tiny\color{codegray},
    stringstyle=\color{codepurple},
    basicstyle=\ttfamily\footnotesize,
    breakatwhitespace=false,         
    breaklines=true,                 
    captionpos=b,                    
    keepspaces=true,                 
    numbers=left,                    
    numbersep=5pt,                  
    showspaces=false,                
    showstringspaces=false,
    showtabs=false,                  
    tabsize=2,
    morekeywords={entry},
}
\DeclareRobustCommand\sampleline[1]{%
  \tikz\draw[#1] (0,0) (0,\the\dimexpr\fontdimen22\textfont2\relax)
  -- (1em,\the\dimexpr\fontdimen22\textfont2\relax);%
}

\def\BibTeX{{\rm B\kern-.05em{\sc i\kern-.025em b}\kern-.08em
    T\kern-.1667em\lower.7ex\hbox{E}\kern-.125emX}}

\newcolumntype{L}[1]{>{\raggedright\arraybackslash}m{#1}}
\newcolumntype{C}[1]{>{\centering\arraybackslash}m{#1}}
\newcolumntype{R}[1]{>{\raggedleft\arraybackslash}m{#1}}

\DeclareCaptionLabelSeparator{periodspace}{.\quad}
\begin{document}

\title{\textsc{Moirai}: Towards Optimal Placement for Distributed Inference on Heterogeneous Devices}
\author{
    \IEEEauthorblockN{Beibei Zhang\IEEEauthorrefmark{2}, Hongwei Zhu\IEEEauthorrefmark{2}, Feng Gao\IEEEauthorrefmark{2}, Zhihui Yang\IEEEauthorrefmark{2}, Sean Xiaoyang Wang\IEEEauthorrefmark{3}}
    \IEEEauthorblockA{\IEEEauthorrefmark{2}Zhejiang Lab, Hangzhou, China}
    \IEEEauthorblockA{\IEEEauthorrefmark{3}Fudan University, Shanghai, China}
}

\maketitle
\input{abstract}
\input{introduction}

\input{backgrounds}
\input{methods}
\input{experiments}

\bibliographystyle{IEEEtran}
\bibliography{references}

\end{document}

%% file: abstract.tex
\begin{abstract}

The escalating size of Deep Neural Networks (DNNs) 
has spurred a growing research interest 
in hosting and serving DNN models across multiple devices. 
A number of studies have been reported to 
partition a DNN model across devices,  
providing device placement solutions. 
The methods appeared in the literature, however, 
either suffer from poor placement performance 
due to the exponential search space 
or miss an optimal placement 
as a consequence of  
the reduced search space with limited heuristics. 
Moreover, these methods have ignored the runtime 
inter-operator optimization of a computation graph when coarsening the graph, 
which degrades the end-to-end inference performance. 
This paper presents \textsc{Moirai} 
that better exploits runtime inter-operator fusion in a model 
to render a coarsened computation graph, 
reducing the search space 
while maintaining the inter-operator 
optimization provided by inference backends.
\textsc{Moirai} also generalizes the device 
placement algorithm from multiple 
perspectives by considering inference constraints and device heterogeneity.
Extensive experimental evaluation with 11 large DNNs 
demonstrates that \textsc{Moirai} outperforms 
the state-of-the-art counterparts, i.e., Placeto, m-SCT, and GETF, 
up to 4.28$\times$ in reduction of the end-to-end inference latency. 
\textsc{Moirai} code is anonymously 
released at \url{https://github.com/moirai-placement/moirai}.

\end{abstract}


%% file: introduction.tex
\section{Introduction}
Recent years have witnessed the prevalence of Deep Neural Networks (DNNs) 
in diverse spectrum of scenarios ranging from object detection~\cite{zhao2019object} 
to text generation~\cite{brown2020language}. 
In these applications, 
researchers increase the DNN model capacity, measured 
by the number of trainable parameters, to achieve 
better model inference accuracy and generalization performance. 
As an example, the state-of-the-art language model Megatron-Turing NLG 
with 530~billion parameters achieves an accuracy score of~87.15\% 
in the task of LAMBADA next word prediction~\cite{smith2022using}. 
However, DNN inference expects considerable memory space 
to store the parameters and intermediate activations, 
which arouses natural concerns about exceeding the memory capacity of a computing device~\cite{tarnawski2020efficient}. 
For instance, GPT-3~\cite{brown2020language} with 175~billion parameters 
requires 350~GB of GPU memory. 
This is far beyond the memory sizes of 
any commercial off-the-shelf GPUs. 

The growing size of DNNs 
necessitates the adoption of heterogeneous resource-constrained 
computing devices to host a large-scale DNN model. 
To accommodate this need, we split a DNN into several sub-models, 
each of which usually consists of continuous operators of a DNN, 
and distribute them across multiple devices. 
The devices execute the disjoint parts of the DNN collectively, 
which is referred to as \textit{model parallelism}~\cite{zheng2022alpa}. 
Fundamentally, model parallelism seeks to map DNN operators to computing devices, commonly termed as \textit{device placement}~\cite{mirhoseini2017device}. 
The objective of the device placement is to minimize the \textit{makespan}, 
which identifies the time interval between a single input and its response. 
In this paper, we focus on providing a device placement solution 
for model parallelism over heterogeneous devices.

Device placement is challenging for two reasons. 
Firstly, allocating operators with precedence constraints 
on separate devices incurs communication overhead 
between the devices due to the \textit{data flow} between devices. 
The overhead varies under distinct device placement schemes.
Secondly, devices possess heterogeneous computing resources. 
Executing an operator on different devices results in different processing time. 
Owing to the discrepancies of model and device configurations, 
device placement poses a huge solution search space 
when the number of operators or devices increases. 
As an example, placing an Inception-v4~\cite{szegedy2017inception} with 490 operators on 2 devices 
has $2^{490}$ different possible placement solutions. 

A direct and intuitive approach to tackle the device placement problem 
is to manually find the DNN partition plan by machine learning experts.
Such an approach might not be favorable for yielding 
an optimal and scalable results. 
Consequently, solutions have been proposed to leverage learning-based methods. 
Particularly, recent studies exploit reinforcement learning techniques, 
which leverage the execution track of similar operators to learn 
where DNN operators should be placed 
in a computer cluster~\cite{mirhoseini2018hierarchical, addanki2019placeto, lan2021accelerated}. 
Unfortunately, learning-based approaches requires a training process 
that takes several hours or even several days to deliver a placement solution for a single DNN~\cite{addanki2019placeto}. 
Additionally, learning-based methods are unable to  
generalize to a different collection of devices. 
The placement solution has to be searched all over again, 
if the DNN is deployed to a different cluster~\cite{jeon2020baechi}. 
The high training overhead and the low generalizability severely 
hinder the widespread application of the learning-based methods. 
To avoid these shortcomings, 
another line of work resorts to algorithmic approaches~\cite{jeon2020baechi, tarnawski2020efficient, hafeez2021towards, zhang2021dynamic}. 
The state-of-the-art algorithmic methods establish cost models 
that reflect end-to-end inference latency 
and propose near-optimal placement solutions through combinatorial optimization. 
According to the combinatorial optimization techniques 
applied to solving the device placement problem, 
we further categorize the status-quo algorithmic approaches into heuristic-based solutions~\cite{jeon2020baechi, zhang2021dynamic, chen2022hare} and exact algorithm-based methods~\cite{tarnawski2020efficient, hafeez2021towards}.


Upon analyzing and experimenting the released implementations of existing algorithmic approaches, 
we observed that the above methods inherently suffer from the following drawbacks:  
(1) \textbf{Solution optimality: }Given model and device configurations, heuristic-based algorithms quickly yield the placement plan. However, the placement result is sub-optimal, leaving ample room for reducing the makespan.
(2) \textbf{Graph optimization: }Machine learning compilers, such as TensorFlow XLA~\cite{abadi2016tensorflow}, TVM~\cite{chen2018tvm}, and PyTorch Glow~\cite{paszke2019pytorch}, represent a DNN as a computation graph with backend optimization to rewrite the graph for reducing the inference makespan. Nevertheless, current algorithmic solutions fail to leverage runtime optimization of a computation graph when coarsening the graph to reduce the solution search space.
(3) \textbf{Device heterogeneity \& Constraints: }Studies revolving around exact algorithms either attempt to produce near-optimal device placement decisions on a small number of homogeneous devices~\cite{tarnawski2020efficient, hafeez2021towards} or do not sufficiently consider the device computation and communication constraints. 

To address the aforementioned limitations, we propose \textsc{Moirai}, 
an algorithmic solution that considers DNN graph optimization 
and caters an optimal solution over heterogeneous devices. 
We embrace two key design considerations to tackle the challenges.
Technically, \textbf{(1)} We leverage a runtime graph optimization, 
\textit{operator fusion}, 
to merge multiple operators into one fused operator 
to reduce the search space for device placement. 
\textbf{(2)} We formalize the problem as 
a \textbf{M}ixed-\textbf{I}nteger \textbf{L}inear \textbf{P}rogramming (MILP) model. 
Unlike the aforesaid approaches, 
in which the device heterogeneity is ignored, 
we craft the MILP to outline computational differences, 
memory constraints, and communication capability of the devices. 
An optimal placement result will then be produced 
by using the optimization solver Gurobi~\cite{gurobi}. 

We conduct \textsc{Moirai} in PyTorch 
and empirically evaluate its efficacy with a total 
of 11 large models over two-device settings. 
We serve the trained Swin-Transformer~\cite{liu2022swin}, 
GPT-3~\cite{brown2020language}, 
and AlphaFold2~\cite{cramer2021alphafold2} 
with the placement plan of \textsc{Moirai} 
to two cluster of devices, each of which consists of 4 GPUs. 
\textsc{Moirai} outperforms the heuristic-based solution up to 1.87$\times$ 
in placement optimality in terms of the reduction of the makespan. 
Compared to the learning-based solution counterparts, 
\textsc{Moirai} reduces the solution generation time from hours to minutes 
while reducing the makespan up to 4.28$\times$. 
Our method is applicable to a range of DNNs 
and may be extended to other situations.

The remainder of the paper is organized as follows. 
First, we introduce background knowledge 
and provide motivation in section~\ref{sec:related-work}. 
Then, we describe the technical details 
of \textsc{Moirai} in section~\ref{sec:proposed-method}. 
Extensive comparisons and evaluations between \textsc{Moirai} 
and its counterparts follow in section~\ref{sec:experiments}. 
Finally, we conclude in section~\ref{sec:conclusion}.

%% file: backgrounds.tex
\section{Background \& Motivation}
\label{sec:related-work}
Before delving into the details of \textsc{Moirai}, 
we provide related works of device placement research. 
We first give a brief overview of distributed deep learning. 
We then present hitherto device placement approaches, 
which are categorized into learning-based solutions and algorithmic methods. 
\begin{figure}[t]
	\centering
	\includegraphics[width=\columnwidth]{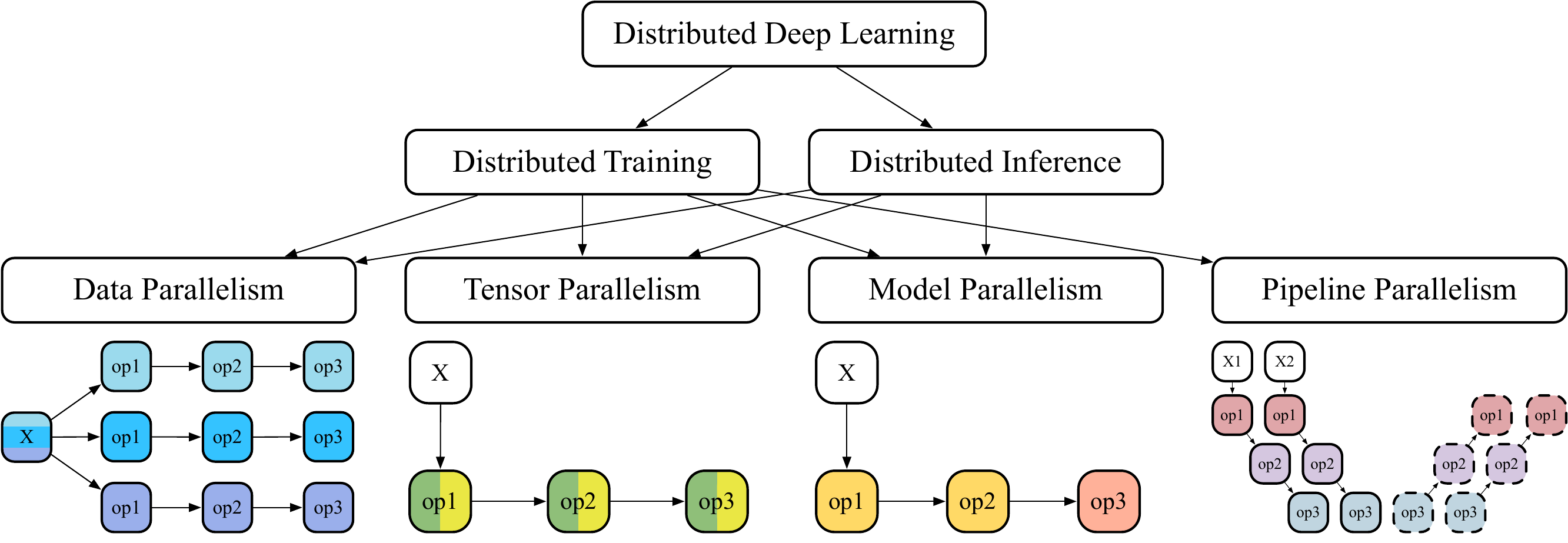}
	\caption{We demonstrate the taxonomy of state-of-the-art distributed deep learning solutions. Under each method, we briefly visualize the layout of the solution silhouettes, where the computation graph includes three operators and each color indicates a distinct device.}
	\label{fig:ddl-classification}
	\vspace{-1em}
\end{figure}


\subsection{Distributed Deep Learning}
In this subsection, we briefly outline popular 
distributed deep learning solutions. 
A DNN life cycle consists of DNN training and DNN inference. 
In the training phase, we split the dataset into multiple mini-batches. 
A mini-batch passes through the operators of a DNN 
in a \textit{forward propagation} phase. 
The output of the forward phase is compared against \textit{ground truth} 
to generate a \textit{loss}, 
which is leveraged to \textit{update} 
the weights of operators in a \textit{backward propagation} phase. 
This process is repeated several times 
to tune the weights until the desired result accuracy is achieved. 
DNN inference performs \textit{forward calculations} using the trained weights. 
Distributed deep learning methods execute the above procedure cooperatively 
on multiple devices aiming to fit the large-scale model 
or obtain computation speedup. 

Fig.~\ref{fig:ddl-classification} depicts the taxonomy of distributed deep learning concepts.
According to the DNN life cycle, 
we categorize distributed deep learning methods 
into distributed training and distributed inference. 
We introduce four major types of distributed machine learning mechanisms, 
namely \textit{data parallelism, tensor parallelism, model parallelism}, and \textit{pipeline parallelism}. 
Data parallelism splits a mini-batch into multiple shards and executes multiple instances of the same
model on decentralized devices with the shards~\cite{rajbhandari2020zero, jiang2020unified}.
Tensor parallelism, also known as \textit{intra-operator model parallelism}, 
partitions an operator along one of its dimension, enabling parallel processing for an operator~\cite{bian2021colossal, hu2022distributed}. 
Model parallelism, commonly referred to as \textit{inter-operator model parallelism},  
divides a DNN into several consecutive sub-models that are placed and computed across multiple devices. 
\textit{Pipeline parallelism} meliorates the model parallelism in the distributed training cycle, 
incorporating the forward phases and the backward stages of several mini-batches to better utilize the idled computation resources~\cite{huang2019gpipe, narayanan2019pipedream, narayanan2021memory}. 
Pipeline parallelism can be deemed as a well-scheduled pipelined model parallelism, which can overlap the computation of different batches. 
A new trend focuses on developing a system that contemplates the mixture of the strategies to achieve the DNN training speedup~\cite{zheng2022alpa, lin2023superscaler}. 
Under each distributed training method, a series of techniques, 
such as \textit{gradient accumulation}, \textit{checkpointing}~\cite{chen2016training}, 
\textit{Parameter Server}~\cite{li2014scaling}, \textit{Ring-AllReduce}~\cite{sergeev2018horovod}, 
are raised to fulfill or optimize its implementation.
Unlikely, we concentrate on inter-operator model parallelism in the distributed inference, 
in pursuit of its optimality, hoping to inspire others.
Other parallelism strategies and technics are orthogonal to our study. 

\begin{figure}[t]
	\centering
	\includegraphics[width=\columnwidth]{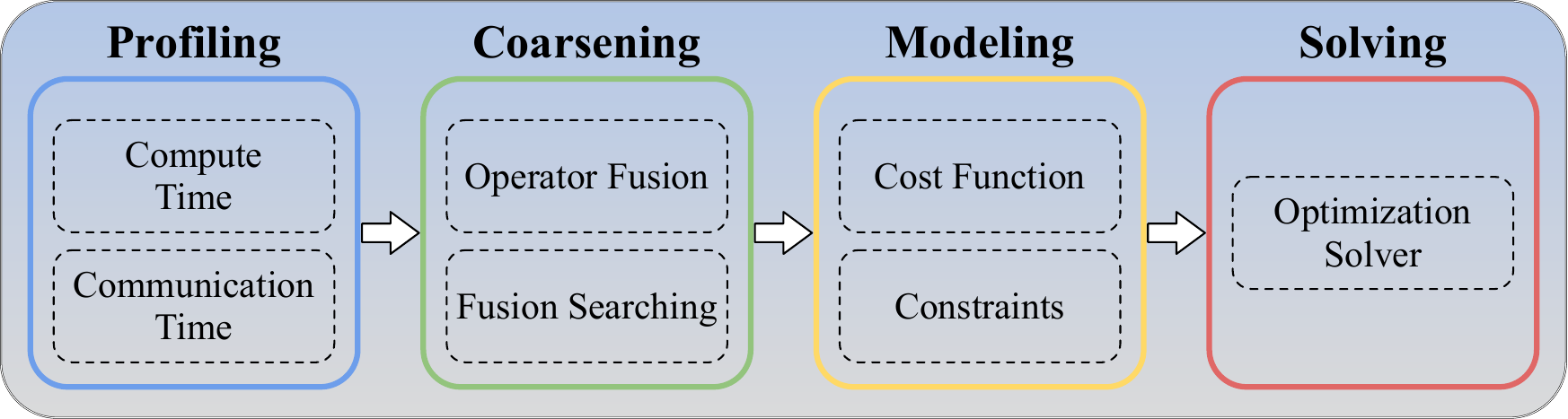}
	\caption{We solve the device placement problem with four steps.}
	\label{fig:overall-proc}
	\vspace{-1em}
\end{figure}

\subsection{Device Placement}

\textbf{Learning-based solutions.} 
Earlier device placement works generally fall within a reinforcement learning paradigm. 
Mirhoseini et al.~\cite{mirhoseini2017device} 
propose to search the placement policy 
for a DNN with reinforcement learning method. 
Precisely, the method carries out a repeated series of 
Monte-Carlo trails with the feedback from real-world servers. 
It adjusts the scheduling strategy 
towards the objective of reducing 
the per-step training time of a DNN until convergence, 
which requires 17-27 hours of 80 to 160 4-GPU servers 
to produce the near-optimal placement. 
To expedite the exorbitant learning process, 
Post~\cite{gao2018post} represents 
device placement as a high-dimensional 
softmax probability distribution. 
Post leverages both proximal policy optimization 
and cross-entropy minimization to achieve a fast convergence. 
In the face of the cues that all previous methods 
can only generate a near-optimal placement 
for a DNN over a set of computing devices, 
Placeto~\cite{addanki2019placeto} encodes computation graphs 
with graph embeddings and offers a policy network 
to disentangle the placement of a family 
of computation graphs with reinforcement learning.

\textbf{Algorithmic methods. }
A variety of heuristics play an indispensable role for device placement problem. 
Inspired by the traditional parallel job scheduling, 
Baechi~\cite{jeon2020baechi} utilizes three heuristics to portray the task precedence, communication overhead, 
and memory constraints of the placement policy, 
which generally requires less than 3 minutes to find a placement strategy. 
To better serve real-world scenarios dominated by heterogeneous devices, 
Hare~\cite{chen2022hare} schedules computation graphs 
with a simple greedy heuristic that always provides higher GPU memory for the next task 
and keeps the latest completed task. 
Directly applying the heuristic methods to the device placement produces unsatisfactory end-to-end latency, 
primarily due to the failure to optimize over the practical computation and communication constraints. 
Accordingly, Pesto~\cite{hafeez2021towards} establish 
an objective function that captures memory and non-overlapping constraints. 
However, Pesto merely considers device placement on two types of devices. 
The proposed method fail to generalize to multiple heterogeneous devices.
GETF~\cite{su2020communication} represents the DNN as a DAG 
and extends the conventional Earliest Time First (ETF) algorithm to 
incorporate related machines. 
GETF establishes a Mixed Integer Linear Programming (MILP) 
model to address the device placement problem. 
Nevertheless, the model neglects to integrate machine-dependent data flow 
communication time as a constraint in the MILP formulation.

\textbf{Graph Coarsening.} 
Given the substantial number of DNN operators, 
device placement algorithms  
encounter a considerable search space 
forfeiting its ability to generate 
placement solutions efficiently and effectively. 
Previous approaches have endeavored to reduce the search space 
by coarsening the computation graph, 
merging operators within the graph with tailored heuristics. 
A pervasive solution is to merge adjacent operators that, 
when combined, do not create cycles~\cite{jeon2020baechi, hafeez2021towards}. 
Recent work considers communication-to-computing ratio 
of the computation graph as a metric to 
guide the grouping of operators~\cite{xu2022celeritas}. 
However, existing approaches fail to leverage the runtime 
optimization of a computation graph.


%% file: methods.tex
\section{Proposed Method}
\label{sec:proposed-method}
Broadly, we layout the overall procedures of \textsc{Moirai} in Fig.~\ref{fig:overall-proc}, 
namely, input profiling, graph coarsening, 
problem modeling, and problem solving. 
We would like to emphasize that 
our approach excels in handling heterogeneous devices 
while maintaining optimal inference latency.
In what follows, we elaborate the comprehensive 
technical details of \textsc{Moirai}. 


\subsection{Assumptions \& Settings}

We enumerate a few practical settings 
and assumptions upon which our algorithm is designed. 
In contrast to the existing approaches, 
we attach substantial importance to the 
heterogeneity of devices, 
mainly in terms of computation, 
memory, and communication differences.

\textbf{DNNs.} The primitive computation unit in a DNN 
is a mathematical operator such as matrix multiplication (\texttt{matmul}) 
or convolution (\texttt{conv}). 
The data flow between operators establishes the 
dependency constraints among the operators. 
A group of operators with precedence constraints constitutes 
a computation graph that describes the DNN inference process. 
The topology of the computation graph is a Directed Acyclic Graph (DAG) 
where vertices represent operators and edges depict precedence constraints. 

\textbf{Devices.} We focus on discussing 
the device placement problem on heterogeneous devices. 
Device heterogeneity is manifested in three folds. 
Firstly, the computation capability of a device is different 
resulting in different processing time of a DNN operator. 
Secondly, the memory capacity of each device is non-uniform, 
suggesting that the number of DNN operators 
that can be hosted by each device varies. 
Thirdly, there are differences in connectivity and bandwidth between devices 
due to their reliance on various network interfaces and protocols.
\begin{figure}[t]
	\centering
	\includegraphics[width=\columnwidth]{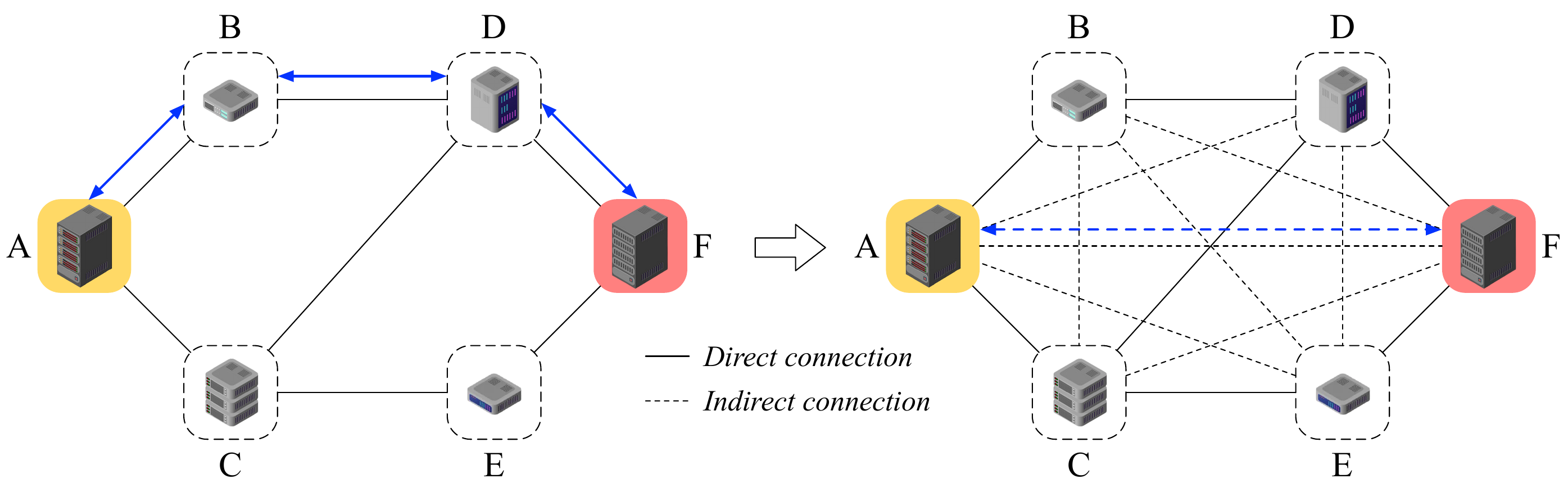}
	\caption{Device $A$ can communicate with device $F$ via the channel $A\rightarrow B\rightarrow D\rightarrow F$, which can be viewed as indirect connection $A$\sampleline{dash pattern=on .3em off .3em on .3em off .3em}$F$. A device cluster (left) can be viewed as a full-mesh (right).}
	\label{fig:full-mesh}
	\vspace{-1em}
\end{figure}

\textbf{Communication. }
We view heterogeneous computers as a collection of connected
devices with one or more addressable network attachments. 
Network attachments may reside at or
above the data link layer and can have various types of
interfaces, such as WiFi, Bluetooth, or even application
defined interfaces. The communication channel between two devices may be provided by a
point-to-point link, a shared broadcast link, or a switched network. 
As illustrated in Fig.~\ref{fig:full-mesh}, 
if two devices in the connected cluster cannot directly communicate with each other, 
they may establish a multi-hop tunnel via internetworking protocols~\cite{perkins2003rfc3561}.
We intensify the connectivity in a connected device cluster with \textit{direct} and \textit{indirect} communication channels, enabling all-to-all communications. 
Therefore, we model the network topology of a connected heterogeneous device cluster as a full-mesh. 
We consider a bidirectional communication network where the availability and the bandwidth of the uplink and downlink are stable.

\subsection{Graph Coarsening}
\begin{figure}[t]
    \centering
	\begin{subfigure}[t]{0.24\columnwidth}
		\centering
		\includegraphics[height=4cm]{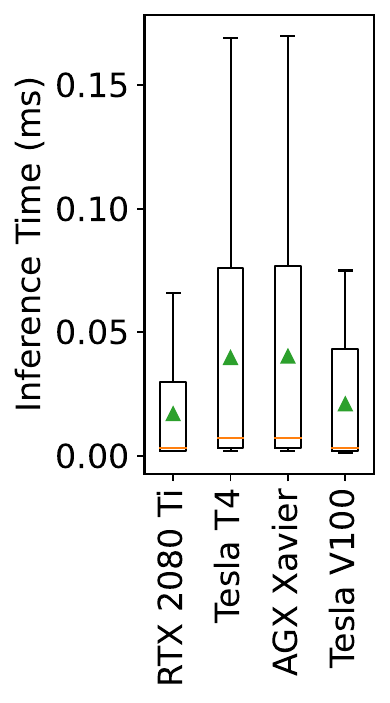}
		\captionsetup{justification=centering}
		\caption{GoogLeNet \\(306 ops)}
		\label{fig:googlenet-ops}
	\end{subfigure}
	\hfill
	\begin{subfigure}[t]{0.24\columnwidth}
		\centering
		\includegraphics[height=4cm]{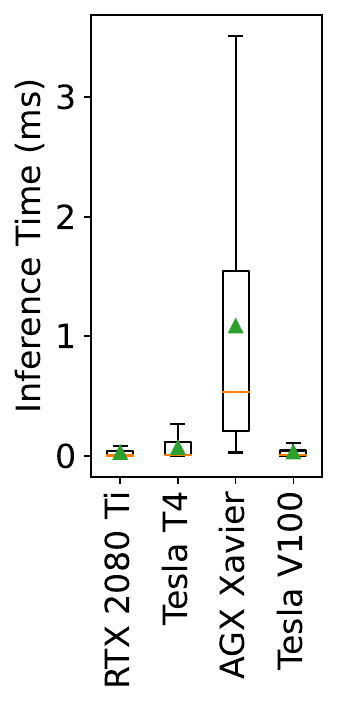}
		\captionsetup{justification=centering}
		\caption{Inception v4 \\(492 ops)}
		\label{fig:inception-ops}
	\end{subfigure}
	\hfill
	\begin{subfigure}[t]{0.24\columnwidth}
		\centering
		\includegraphics[height=4cm]{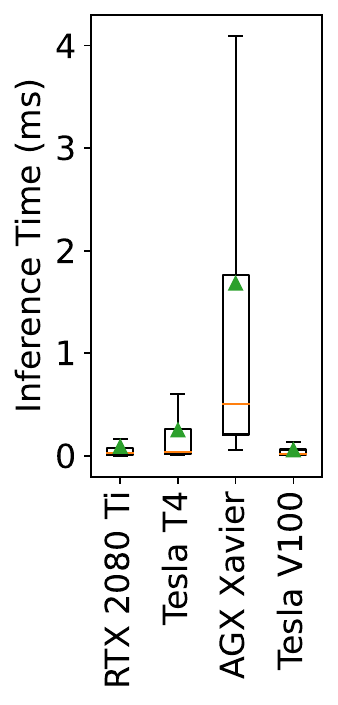}
		\captionsetup{justification=centering}
		\caption{YOLOv5x \\(420 ops)}
		\label{fig:yolov5x-ops}
	\end{subfigure}
	\hfill
	\begin{subfigure}[t]{0.24\columnwidth}
		\centering
		\includegraphics[height=4cm]{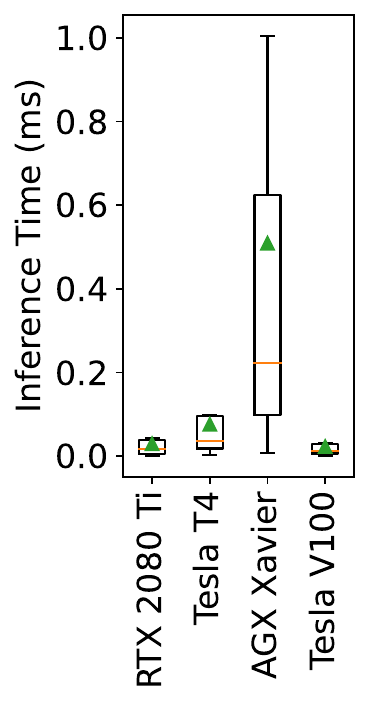}
		\captionsetup{justification=centering}
		\caption{Transformer \\(1375 ops)}
		\label{fig:transformer-ops}
	\end{subfigure}
	\caption{Inference time distribution of operators in four models on four devices.`-' indicates median and `$\vartriangle$' presents mean.}
	\label{fig:operator-proc-time}
	\vspace{-1em}
\end{figure}

We profile operator processing time of four widely employed DNNs on four devices
and show their processing time 
distribution in Fig.~\ref{fig:operator-proc-time}.
We note that modern DNNs typically consist of a number of 
operators with short computation time, 
increasing the difficulty to address the device placement problem. 
To reduce the solution search space, we coarsen the computation graph 
by grouping the closely coupled operators 
to promote the device placement algorithm. 

\begin{figure}[t]
	\centering
	\includegraphics[width=0.85\columnwidth]{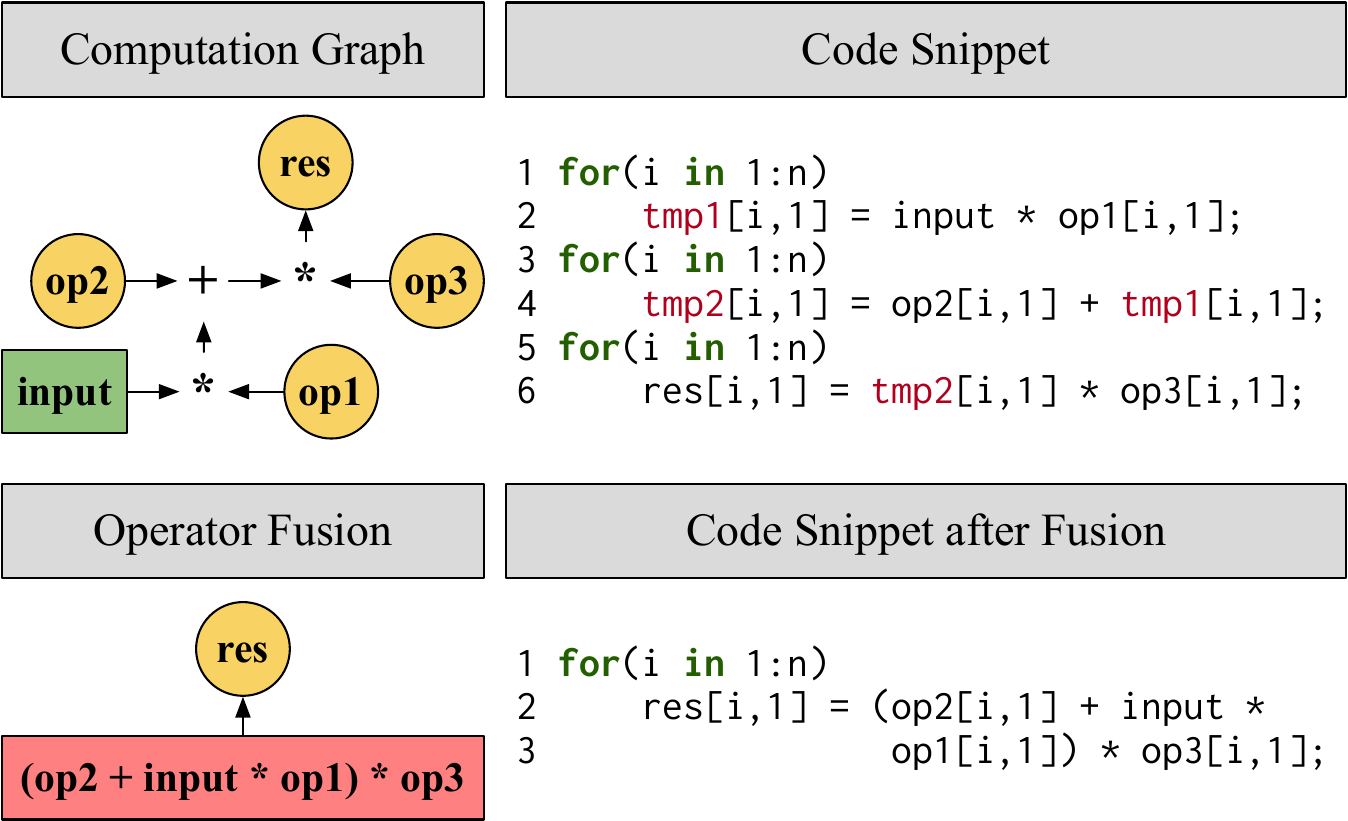}
	\caption{Operator op1, op2, and op3 are fused by the backend compiler to avoid temporary results.}
	\label{fig:operator-fusion}
	\vspace{-1em}
\end{figure}
\textbf{Operator fusion.} An inference backend, such as Eigen~\cite{eigenweb} and NNPack~\cite{nnpack}, combines operators that satisfy certain operator types and connections into a single fused operator, 
referred to as \textit{operator fusion}.
Operator fusion avoids storing intermediate results in memory, reducing frequent memory accesses. 
The memory access latency is often orders of magnitude greater than the computation time, 
and thus operator fusion extensively speeds up the DNN inference.
We show an example in Fig.~\ref{fig:operator-fusion}, where operator \texttt{op1}, \texttt{op2}, and \texttt{op3} are fused into one operator to produce the final result \texttt{res}. 
Fusing \texttt{op1}, \texttt{op2}, and \texttt{op3} avoids the necessity to store and access 
the temporary results of calculations against \texttt{op1} and \texttt{op2}.
Upon the above observations, we intrinsically believe 
that the fused operators should be placed together when we effectuate graph coarsening.
\begin{table}[t]
    \centering
    \setlength\extrarowheight{1pt}
    \caption{We take several operator fusion rules provided by Eigen on a GPU kernel as examples.}
    \resizebox{\columnwidth}{!}{
	\begin{tabular}{|| C{2em} |  C{12em} | C{12em} ||}
		\hline
		\textbf{ID} & \textbf{Operators} & \textbf{Fused Operator} \\ 
		\hline
		1 & \texttt{conv, bn} & \texttt{conv} $\circ$ \texttt{bn}\\
		\hline
		2 & \texttt{conv, bn, relu} & \texttt{conv} $\circ$ \texttt{bn} $\circ$ \texttt{relu}\\ 
		\hline
		3 & \texttt{conv, bn, add, relu} & \texttt{conv} $\circ$ \texttt{bn} $\circ$ \texttt{add} $\circ$ \texttt{relu} \\
		\hline
	\end{tabular}
}
    \label{tab:fusion-rules}
\end{table}

\begin{figure}[t]
	\centering
	\begin{subfigure}[t]{0.12\textwidth}
		\centering
		\includegraphics[height=1.3cm]{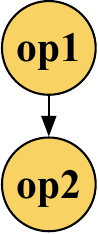}
		\caption{Direct connection.}
		\label{fig:direct-connection}
	\end{subfigure}
	\hfill
	\begin{subfigure}[t]{0.12\textwidth}
		\centering
		\includegraphics[height=1.3cm]{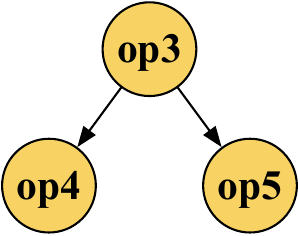}
		\caption{Multi-outputs.}
		\label{fig:multi-outputs}
	\end{subfigure}
	\hfill
	\begin{subfigure}[t]{0.12\textwidth}
		\centering
		\includegraphics[height=1.3cm]{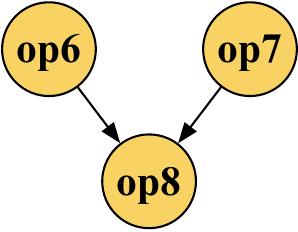}
		\caption{Multi-inputs.}
		\label{fig:multi-inputs}
	\end{subfigure}
	\caption{Three types of operator connections.}
	\label{fig:connection-types}
\end{figure}

An inference backend uses, for example, the \textit{fusion rules} shown in TABLE~\ref{tab:fusion-rules},  
to define which operators in a DNN should be fused~\cite{boehm2018optimizing}. 
A fusion rule contains a sequence of ordered operator types, each of which is a string. 
Fusion rules can be obtained from the design specifications 
of inference backends.
Typically, the connections of DNN operators can be categorized 
into three types in a computation graph, 
which are illustrated in Fig.~\ref{fig:connection-types}, namely \textit{direct connection}, \textit{multi-outputs}, and \textit{multi-inputs}. 
As pointed out by~\cite{zhang2021nn}, 
given a fusion rule, only fusing the operators with direct connection or multi-inputs connection 
can optimize the inference speed. 

\textbf{Fusion searching.} Given a DNN, \textsc{Moirai} coarsens its computation graph by grouping operators based on fusion rules. 
We depict the operators as a set of vertices $\{v_1, v_2, \dots, v_n\}$ in a graph, 
where $n$ is the number of operators in the DNN and $v_i$ is the $i$-th DNN operator. For any two vertices $v_i$ and $v_j$ in the graph, we introduce a directed edge $(i, j)$ in the graph if and only if there is a data transfer from operator $i$ to $j$. With the above settings, we untangle the given DNN by the following DAG
\begin{equation}
\label{eqn:dag}
	\mathcal{G} = (\mathcal{V}, \mathcal{E}), 
\end{equation}
where $\mathcal{V}=\{v_1, v_2, \dots, v_n\}$ and $\mathcal{E} \subseteq \mathcal{V}\times \mathcal{V}$. 
We theoretically define fusion rules as a set $\mathcal{R}=\{r_1, r_2, \dots, r_m\}$, 
where $m$ is the ID of a rule and the $i$-th rule $r_i\in \mathcal{R}$ is an ordered list with operator types as elements. 
The element order in $r_i$ suggests the precedence constraints of the fused operators. 
Mathematically, \textsc{Moirai} engages in validating and grouping vertices in $\mathcal{G}$ pursuant to a set of ordered list $\mathcal{R}$, which is described in Algorithm~\ref{alg:fusion-coarsen}. 

\begin{figure}[t]
	\centering
	\begin{subfigure}[t]{0.24\textwidth}
		\centering
		\includegraphics[height=1.2cm]{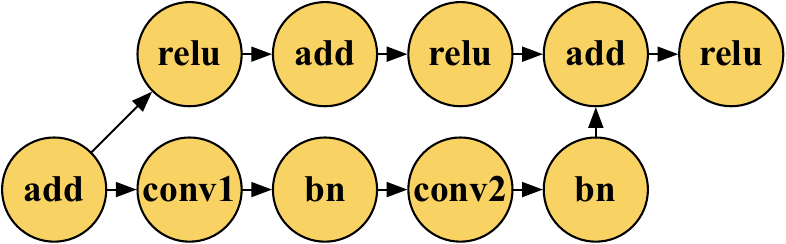}
		\caption{DNN computation graph.}
		\label{fig:gcof-example-1}
	\end{subfigure}
	\hfill
	\begin{subfigure}[t]{0.24\textwidth}
		\centering
		\includegraphics[height=1.3cm]{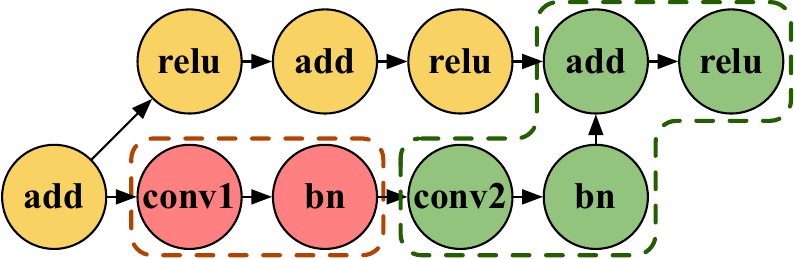}
		\caption{Operator grouping.}
		\label{fig:gcof-example-2}
	\end{subfigure}
	\caption{$\mathtt{GCOF}()$ groups the DNN operators according to the TABLE~\ref{tab:fusion-rules}.}
	\label{fig:gcof-example}
	\vspace{-1em}
\end{figure}

We elaborate the algorithm with an example in Fig.~\ref{fig:gcof-example} where we follow the fusion rules provided by TABLE~\ref{tab:fusion-rules}. 
$\mathtt{GCOF}()$ traverses the graph in depth-first order from the root vertex of operator type \texttt{add}. 
The function first navigates to the upper branch of the graph. 
Although the first \texttt{add}, \texttt{relu} vertex pair 
is certified to conforming partial $r_3$ rule via the function $\mathtt{is\_sub\_rule}()$, 
the operator connection denoted by the edge between \texttt{add} and \texttt{relu} vertex
is essentially part of the multi-output connection of the operator \texttt{add}, 
which is examined by function $\mathtt{is\_valid\_conn}()$.
Therefore, the first pair of \texttt{add}, \texttt{relu} should not be fused. 
The function then moves forward and binds the other two \texttt{add}, \texttt{relu} vertex pair on the upper branch, 
generating two new vertices with the $\mathtt{bound}$ tag and the \texttt{add}~$\circ$~\texttt{relu} type. 
In the lower branch of the DNN, $\mathtt{GCOF}()$ fuses \texttt{conv1} and \texttt{bn}, 
which comply with rule $r_1$ and the direct connection.
Likewise, \texttt{conv2} and \texttt{bn} are fused. 
Next, $\mathtt{GCOF}()$ merges the operator \texttt{conv2} $\circ$ \texttt{bn} and operator \texttt{add} $\circ$ \texttt{relu} at the end of the upper branch in accordance with rule $r_3$ and multi-inputs connection.
Function $\mathtt{unbind}()$ releases the operators with the $\mathtt{bound}$ tag, which is the second \texttt{add}, \texttt{relu} pair on the upper branch in our case. The output of Algorithm~\ref{alg:fusion-coarsen} is a coarsened graph $G$ of DAG topology.
The time complexity of Algorithm~\ref{alg:fusion-coarsen} is $O(\mathcal{V}+\mathcal{E})$. 


\begin{algorithm}[t]
	\caption{Graph Coarsening with Operator Fusion: $\mathtt{GCOF}()$}
	\label{alg:fusion-coarsen}
	\KwData{DAG: $\mathcal{G=(V, L)}$ \; \hspace{0.83cm} Fusion rules: $\mathcal{R}$.}
	\KwResult{Coarsened graph $G$ by operator fusion.}
	\SetKwProg{Fn}{function}{:}{}
	\Fn{$\mathtt{fuse}(v_{pred}, v_{succ})$}{
	    $v_{new}\leftarrow$ initiate a new operator\;
	    $v_{new}.in \leftarrow v_{pred}.in \cup v_{succ}.in - v_{pred}$\;
	    $v_{new}.out \leftarrow v_{pred}.out \cup v_{succ}.out - v_{succ}$\;
	    $v_{new}.type \leftarrow v_{pred}.type \circ v_{succ}.type$\;
	    $v_{new}.tag \leftarrow \mathtt{fused}$\;
	    return $v_{new}$\;
	}
	\BlankLine
	\Fn{$\mathtt{bind}(v_{pred}, v_{succ})$}{
	    $v_{new} \leftarrow \mathtt{fuse}(v_{pred}, v_{succ})$\;
	    $v_{new}.tag \leftarrow \mathtt{bound}$\;
	    return $v_{new}$\;
	}
	\BlankLine
	\Fn{$\mathtt{dfs}(v_{pred})$}{
		\ForEach{$v_{succ}$ in $v_{pred}.out$}{
			\uIf{$\mathtt{is\_rule}(v_{pred}, v_{succ}, \mathcal{R})$ and $\mathtt{is\_valid\_conn}(v_{pred}, v_{succ})$}{
		        $v_{next}\leftarrow \mathtt{fuse}(v_{pred}, v_{succ})$\;
		        add $v_{next}$ in $\mathcal{G}$\;
		        remove $v_{pred}$, $v_{succ}$ in $\mathcal{G}$
			}
			\uElseIf{$\mathtt{is\_sub\_rule}(v_{pred}, v_{succ}, \mathcal{R})$ and $\mathtt{is\_valid\_conn}(v_{pred}, v_{succ})$}{
				$v_{next}\leftarrow \mathtt{bind}(v_{pred}, v_{succ})$\;
			}
			\Else{
			    $v_{next}\leftarrow v_{succ}$\;
			}
		}
		$\mathtt{dfs}(v_{next})$\;
	}
	\BlankLine
	\Fn{$\mathtt{unbind}(\mathcal{G})$}{
	    release all the operators with the $\mathtt{bound}$ tag in $\mathcal{G}$\;
	}
	\BlankLine
	$v_{pred}\leftarrow$ initiate traversal with the root vertex of $\mathcal{G}$\;
	$\mathtt{dfs}(v_{pred})$\;
	$G\leftarrow \mathtt{unbind}(\mathcal{G})$\;
\end{algorithm}

\subsection{Input Profiling}

Our method takes compute time of 
each operator, transmission time of data flow, precedence relation among operators, 
and configurations of devices as its inputs. 
The operator dependency is manifested by the computation graph 
and the configurations of devices can be 
acquired by querying operating system interfaces, 
whereas the operator processing and 
the data transmission time requires proper analysis.

\textbf{Compute time. } 
Scrutinizing the existing research on measuring 
DNN operator processing time, 
we notice that there are fundamentally 
three commonly employed methods to profile the operator processing time, 
that is \textit{manual testing}, \textit{operational intensity}, 
and \textit{prediction model}. 
Though manual testing reveals the actual operator execution time, 
it is labor intensive to approach the data. 
Operational intensity~\cite{williams2009roofline} theoretically evaluates the task computation latency. 
However, several factors other than 
memory-bound and compute-bound affect 
the actual operator processing time. 
To balance the operator processing time accuracy and its availability, 
\textsc{Moirai} chooses to estimate the compute time following the ideas in~\cite{geoffrey2021habitat}. 

\textbf{Communication time. }
Communication time between two devices 
amounts to the ratio of the data flow size and the communication bandwidth. 
In an indirect communication channel, 
the bandwidth of a multi-hop path depends on the minimum bandwidth on the path. 
Take the device cluster in Fig.~\ref{fig:full-mesh} as an instance, 
suppose that the bandwidth of link $A-B$ and link $B-D$ is 10MB/s and 5MB/s, 
transmitting a 100MB data requires 20s. 

\subsection{Problem Modeling}
\begin{figure}[t]
	\centering
	\includegraphics[width=0.85\columnwidth]{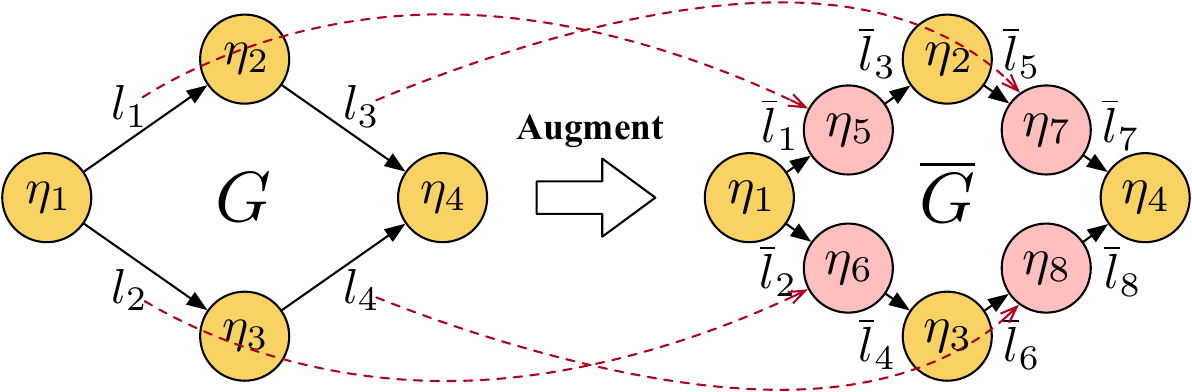}
	\caption{We convert the links into new nodes that hold the same weight, resulting in a augmented DAG $\overline{G}$. Additional direct links without weights are inserted among the original nodes and the newly added nodes to maintain the node precedence and the graph topology. }
	\label{fig:dag-augment}
\end{figure}
Up to this point, we have presented how to obtain a
coarsened graph to reduce the solution search space 
and prepared the necessary inputs for our algorithm. 
Next, we introduce our MILP model seeking to 
portray the inter-operator model parallelism.

\begin{table}[t]
    \centering
    \setlength\extrarowheight{1.5pt}
    \caption{Nomenclature.}
	\begin{tabular}{|| L{6em} |  L{22em} ||}
		\hline
		\textbf{Notations} & \textbf{Descriptions} \\ 
		\hline
		\multicolumn{2}{||c||}{\textit{Parameters}} \\
		\hline
		$G=(N, L)$ & The DAG of the coarsened computation graph. \\
		\hline
		$\overline{G}=(\overline{N}, \overline{L})$ & The augmented DAG of $G$, where links are converted to nodes. \\ 
		\hline
		$\mathtt{Succ}(i)$ & A set of direct and indirect successors of node $\eta_i$ in $G$, where  $\eta_i\in N$ . \\
		\hline
		$\overline{\mathtt{Succ}}(i)$ & A set of direct and indirect successors of node ${\eta}_i$ in $\overline{G}$, where ${\eta}_i\in \overline{N}$. \\
		\hline
		$M^s, M^l, M^r$ & Three large numbers where $M^s\gg 0$, $M^l\gg 0$, $M^r\gg 0$. \\
		\hline
		\multicolumn{2}{||c||}{\textit{Variables}} \\
		\hline
		$k$ & $k\in K$ is the index of a device, which specifies the device.\\
		\hline
		$p_{ik}$ & The processing time of $i$-th operator on the $k$-th device, where $\eta_i\in N$. \\
		\hline
		$p_{qk'k''}^{\text{comm}}$ & The transmission time of data flow $q$ from device $k'$ to device $k''$, where ${\eta}_q \in \overline{N}-N$ and $k',k'' \in K$. \\
		\hline
		$m_i$ & The memory footprint of $i$-th operator. \\
		\hline
		$Mem_k$ & The memory size of $k$-th device. \\
		\hline
		$S_i$ & $S_i\in \mathbb{R}^+$, expresses the start time of task $i$ where ${\eta}_i \in \overline{N}$.  \\
		\hline
		$C_i$ & $C_i\in \mathbb{R^+}$, expresses the complete time of task $i$ where ${\eta}_i \in \overline{N}$.  \\
		\hline
		\multicolumn{2}{||c||}{\textit{Indicators}} \\
		\hline
		 $x_{ik}$ & $x_{ik}\in \{0,1\}$, $x_{ik}=1$ indicates placing the task denoted by $\eta_i\in N$ to device $k$, otherwise $x_{ik}=0$.  \\
		\hline
		 $z_{q}$ & $z_{q}\in \{0,1\}$, $z_q=1$ indicates a non-zero transmission time of the data flow $q$, where ${\eta}_q \in \overline{N}-N$. Otherwise $z_q=0$.  \\
		\hline
		$u_{qk'k''}$ & $u_{qk'k''} \in \{0,1\}$, $u_{qk'k''}=1$ indicates the selection of communication channel from device $k'$ to $k''$ to transmit the data flow $q$, where ${\eta}_q \in \overline{N}-N$, $k',k'' \in K$ and $k' \neq k''$.\\
		\hline
	\end{tabular}
    \label{tab:notations}
\end{table}

\textbf{DAG representation.} Given a coarsened computation graph, 
we refer to the set of $\alpha$ operators in the coarsened graph 
as a set of nodes $N=\{\eta_i\}_{i=1}^{\alpha}$. 
Further, we depict the data flow among the operators as a set of links $L=\{l_i\}_{i=1}^{\beta}$.
Thus, we model the coarsened computation graph as a DAG 
\begin{equation}
\label{eqn:coarsened-dag}
G=(N, L), 
\end{equation}
where $|N|=\alpha$, $|L|=\beta$, and $L\subseteq N\times N$. 
The node weight represents operator processing time on devices 
and link weight exhibits data transmission time over a communication network.
To facilitate our design, we augment the DAG $G$ 
by altering the links into a set of new nodes. 
Subsequently, the weight of the link is directly applied to the new node corresponding to it. 
We denote the augmented DAG of $G$ by 
\begin{equation}
\label{eqn:augmented-dag}
\overline{G}=(\overline{N}, \overline{L}), 
\end{equation}
where $\overline{N}=\{\eta_i\}_{i=1}^{\alpha+\beta}$ 
and $\overline{L}=\{\overline{l}_i\}_{i=1}^{2\beta}$. 
$\overline{l}_i \in \overline{L}$ can be indicated by the index of its two end points (e.g., $\overline{l}_1=(1, 5)$ in Fig.~\ref{fig:dag-augment}).
Given the DAG $G$ of a DNN, 
we conveniently leverage $\eta_i \in \overline{N}$ to outline both the data flow and the operator.
To identify data flow and operators respectively, we refer to a data flow as $\eta_i \in \overline{N}-N$ and an operator as $\eta_i \in N$.
An example of the relationship between $G$ and $\overline{G}$ is shown in Fig.~\ref{fig:dag-augment}.

\textbf{MILP model.} We summarize key notations of the model in TABLE~\ref{tab:notations}. 
Presented with the DNN computation graph termed as $G=(N, L)$, 
the device placement for operators in the DNN is achieved by solving the following MILP.
\begin{align}
\mathop{\textrm{minimize}}\ \ \  & \max_{i \in N} C_i \label{eqn:ilp}, \\
\textrm{subject to} \ \ \ 
       & C_i \leq S_j, \hspace{2.9em}\forall {\eta}_i \in \overline{N}, \forall {\eta}_j \in \overline{\mathtt{Succ}}(i), \tag{\ref{eqn:ilp}a}  \label{eqn:ilp-a} \\
       & C_i = S_i + \sum_{k\in K}p_{ik}x_{ik}, \hspace{1.39em}\forall \eta_i \in N, \tag{\ref{eqn:ilp}b}  \label{eqn:ilp-b} \\
       & \sum_{k\in K} x_{ik} = 1, \hspace{5.23em}\forall \eta_i \in N, \tag{\ref{eqn:ilp}c}  \label{eqn:ilp-c} \\
       & \textrm{Memory constraints}, \tag{\ref{eqn:ilp}d}  \label{eqn:ilp-d} \\
       & \textrm{Non-overlapping constraints}, \tag{\ref{eqn:ilp}e}  \label{eqn:ilp-e} \\
       & \textrm{Communication constraints},  \tag{\ref{eqn:ilp}f}  \label{eqn:ilp-f} \\
       & \textrm{Congestion control}. \tag{\ref{eqn:ilp}g}  \label{eqn:ilp-g}
\end{align}

Equation~\eqref{eqn:ilp} represents the completion time of the last operator that ends the computation, 
which amounts to the end-to-end inference latency of the entire DNN 
when the inference starts at time 0. 
The objective function~\eqref{eqn:ilp} is optimized subject to the constraints 
from equation~\eqref{eqn:ilp-a} to~\eqref{eqn:ilp-g}. 
The data flow of a DNN, defined by the input and output of its operators, 
naturally establishes the execution precedence relationships of the operators. 
That is, the successor of an operator can only be processed 
after the completion of the current operator. 
Moreover, the output of the operator can only be 
transmitted after it is produced. 
We cast such operator processing and data transmission 
dependencies with equation~\eqref{eqn:ilp-a}. 
Equation~\eqref{eqn:ilp-b} bridges the relation between the start time and end time of processing an operator. 
We ensure that each operator is assigned to only one device by equation~\eqref{eqn:ilp-c}.

Constraints~\eqref{eqn:ilp-d} to~\eqref{eqn:ilp-g} account for the heterogeneity of devices. 

(1) \textit{Memory constraints.} Conventionally, the cumulative memory footprint of the operators allocated to a device 
should not surpasses the memory size of the device, known as memory constraints. 
We describe the memory constraints~\eqref{eqn:ilp-d} with 
\begin{equation}
    \underbrace{\sum_{\eta_i \in N} m_ix_{ik}}_{\substack{\text{Total memory of} \\ \text{operators on device } k}} \leq Mem_k, \forall k \in K, 
    \label{eqn:mc}
\end{equation}
to avoid the out of memory (OOM) error. 
For each device, we obtain the memory footprint of each operator through the APIs (e.g., $\mathtt{torch.profiler}$) 
and constrain the total memory of the operators placed on a device not to exceed the memory capacity of the device. 

(2) \textit{Non-overlapping constraints.} By default, inference frameworks, 
such as PyTorch and TensorFlow, execute operators placed on the same device sequentially. 
Therefore, for any two operators assigned to the same device, we ensure that the their processing time does not overlap. 
Equation~\eqref{eqn:ilp-a} maintains the order of execution for the two operators with precedence relationship. 
Given the two operators $i$ and $j$ without precedence constraints,
we mathematically express the non-overlapping condition with 
\begin{flalign}\label{eqn:noc}
    \begin{cases}
        S_i \geq C_j- M^s\delta_{ij} - M^l(2-x_{ik}-x_{jk}), \\
        S_j \geq C_i- M^s(1-\delta_{ij}) - M^l(2-x_{ik}-x_{jk}), \\
        i\neq j, \\
        \forall \eta_i,\eta_j \in N, \\
        \eta_i\not\in \mathtt{Succ}(j) \text{ and } \eta_j \not\in \mathtt{Succ}(i),\\
        \forall k \in K,
    \end{cases}
\end{flalign} 
where $\delta_{ij} \in \{0,1\}$ is a 0-1 indicator variable. 
If two operators $i$ and $j$ are placed on the device $k$, which is termed as $x_{ik}=x_{jk}=1$, 
the inequalities in constraints~\eqref{eqn:noc} are 
\begin{flalign}
	\begin{cases}
    S_i &\geq C_j- M^s\delta_{ij}, \nonumber\\
    S_j &\geq C_i- M^s(1-\delta_{ij}), \nonumber
    \end{cases}
\end{flalign}
in which both inequalities hold when $\delta_{ij}$ takes different values. For every two operators of a DNN, we apply the constraints in \eqref{eqn:noc}.

(3) \textit{Communication constraints.} Naturally, when two adjacent operators are placed on two distinct devices, 
a communication overhead is incurred by the data flow between the operators. We formalize the communication overhead with the indicator $z_q$. Moreover, our model depicts bandwidth difference of the uplink bandwidth and downlink bandwidth between two devices. We capture the selection of channels between two devices with the indicator $u_{qk'k''}$. 
\begin{equation}\label{eqn:cc}
		\begin{dcases}\left.
			\begin{aligned}
				z_q &\leq 2 - x_{ik} - x_{jk} \\
				z_q &\geq x_{ik} - x_{jk} \\
				z_q &\geq x_{jk} - x_{ik}
			\end{aligned} \right \}, 	\begin{aligned}\forall q \in \overline{N}-N, \\(i,q),(q,j) \in \overline{L},\end{aligned} & \\
		    \left.
		    \begin{aligned}
			    &\sum_{k' \in K}\sum_{k'' \in K}u_{qk'k''}=z_q\\
				&u_{qk'k''} \geq x_{ik'} + x_{jk''} -1 \\  
				&C_q = S_q + \underbrace{\sum_{k' \in K}\sum_{k'' \in K}u_{qk'k''} \cdot p_{qk'k''}^{\text{comm}}}_{\substack{\text{Transmission time of data flow } q\\ \text{ over the channel } k'\rightarrow k''.}}
			\end{aligned} \right \}, \begin{aligned}\forall q \in \overline{N}-N, \\(i,q),(q,j) \in \overline{L}, \\ \forall k’,k'' \in K, \\ k'\neq k''. \end{aligned} &
		\end{dcases}
\end{equation}
Given two contiguous operators $i$ and $j$, if only one of them is placed on device $k$, which implies that there is a communication overhead for data flow $q$ , $z_q=1$ is enforced by the first three inequalities in constraints~\eqref{eqn:cc}. Otherwise, $z_q=0$. The channel where data flow $q$ is transmitted is indicated by $u_{qk'k''}$. If the transfer of data flow $q$ exists, which implies $z_q=1$, the fourth equation ensures that the communication task $q$ selects at most one communication channel $k'\rightarrow k''$ for transmission. The fifth constraint indicates that the communication task $q$ selects the channel $k'\rightarrow k''$ for transmission only when task $i$, task $j$ are deployed on device $k'$ and device $k''$ respectively (i.e. $x_{ik'} = x_{jk''}=1$). 
The last equation in~\eqref{eqn:cc} bridges the start and end time of transmitting the data flow $q$.

(4) \textit{Congestion control.} Lastly, we address the contention of data transmission, when there are multiple outputs waiting to be transferred on the same communication channel. 
Given a device $k$ and two pairs of adjacent operators $a,b$ and $c,d$ where $(a,q), (q,b), (c,r), (r,d) \in \overline{L}$, 
the congestion happens when $x_{ak}=1, x_{bk}=0$ and $x_{ck}=1,x_{dk}=0$. 
In other words, two communication operations should not be processing simultaneously on the same channel. 
In other words, either $S_q \geq C_r$ or $S_r\geq C_q$ holds. 
Formally, the congestion control can be casted as
\begin{align}\label{eqn:ccl}
    \begin{cases}
        S_q \geq C_r - M^s\delta_{qr} - M^l(2-z_q-z_r) \\
        \phantom{S_q \geq C_r}+ M^r(x_{ak}+x_{ck}-x_{bk}-x_{dk}-2), \\
        S_r \geq C_q - M^s(1-\delta_{qr}) - M^l(2-z_q-z_r) \\
        \phantom{S_r \geq C_q}+ M^r(x_{ak}+x_{ck}-x_{bk}-x_{dk}-2), \\
        S_q \geq C_r - M^s\delta_{qr} - M^l(2-z_q-z_r) \\
        \phantom{S_q \geq C_r}+ M^r(x_{bk}+x_{dk}-x_{ak}-x_{ck}-2), \\
        S_r \geq C_q - M^s(1-\delta_{qr}) - M^l(2-z_q-z_r) \\
        \phantom{S_r \geq C_q}+ M^r(x_{bk}+x_{dk}-x_{ak}-x_{ck}-2), \\
        q\neq r,\\
        \forall \eta_q, \eta_r \in \overline{N}-N, \\
        \eta_q \not\in \overline{\mathtt{Succ}}(r) \text{ and } \eta_r \not\in \overline{\mathtt{Succ}}(q), \\
        (a,q), (q,b), (c,r), (r,d) \in \overline{L}, \\
        \forall k \in K,
    \end{cases}
\end{align}
where $\delta_{qr}\in \{0,1\}$ is a 0-1 indicator variable.
For multiple transmission tasks on the same communication channel, we bound every two communication tasks by the constraints~\eqref{eqn:ccl}.

\begin{table*}[th]
	\centering
	\setlength\extrarowheight{1pt}
	\caption{Testbed configurations of two experiment scenarios.}
	\resizebox{\textwidth}{!}{
	\begin{tabular}{cccccccc}
		\toprule
		\multirow{2}{*}{\textbf{Scenario}} & \multirow{2}{*}{\textbf{Device}} & \multirow{2}{*}{\textbf{Memory (GB)}} &\multirow{2}{*}{\textbf{Network Interface}} & \multicolumn{4}{c}{\textbf{Average Network Bandwidth (Gbps)}}\\ 
		& & & & Device A & Device B & Device C & Device D \\
		\hline
		\multirow{4}{*}{\shortstack[c]{Inter-server}}& A: NVIDIA GeForce RTX 2080 Ti & 11 &\multirow{4}{*}{InfiniBand}  & N.A. & 44.26 & 32.92 &  44.28 \\
		& B: NVIDIA Tesla T4 & 16 & & 42.39 & N.A. & 35.32 & 44.51\\
		& C: NVIDIA Tesla P4 & 8	 & & 33.2 &  35.31 & N.A. & 32.95\\
		& D: NVIDIA RTX 3060 Ti & 8	 & & 42.08 & 43.22 & 33.28 & N.A.\\
		\hline
		\multirow{4}{*}{\shortstack[c]{Intra-server}}& A: NVIDIA Tesla V100 & 32 &\multirow{4}{*}{NVLink + NVSwitch} & N.A. & 1170.04 & 626.10 & 610.56\\
		& B: NVIDIA Tesla V100 & 32 & & 1148.16 & N.A. & 618.98 & 581.09 \\
		& C: NVIDIA Tesla P100 & 16 & & 630.43 & 609.82 & N.A. & 571.96\\
		& D: NVIDIA Tesla P100 & 16 & & 622.67 & 575.08 & 581.35 & N.A.\\
		\bottomrule
	\end{tabular}
    }
	\label{tab:device-config}
\end{table*}

\begin{table*}[th]
	\centering
	\setlength\extrarowheight{1.5pt}
	\caption{Model architecture with increasing number of parameters. M: million. B: billion.}
	\resizebox{\textwidth}{!}{
	\begin{tabular}{c c c c c c c c c}
		\toprule
		\textbf{Model} & \textbf{Parameters} & \textbf{Layer Number} & \textbf{Hidden Size} & \textbf{Head Number} & \textbf{N.O. Operators in Original Graph} & \textbf{N.O. Operators in Coarsened Graph}\vspace{4pt}\\
		\hline
		Swin-Transformer~\cite{liu2022swin} & \{1.8B, 6.6B, 13B\} & \{32, 48, 56\} & \{512, 768, 1024\} & \{16, 24, 32\} & \{6496, 14352, 22120\} & \{5204, 11512, 17947\}\\ 
        GPT-3~\cite{brown2020language} & \{330M, 1.3B, 2.7B, 13B\} & \{24, 32, 32, 40\} & \{1024, 2048, 2560, 5120\} & \{16, 32, 32, 40\} & \{4872, 9480, 12640, 19640\} & \{3682, 7308, 9825, 15283\}\\ 
        AlphaFold2~\cite{AlphaFold2021} & \{87M, 930M, 2.4B, 3.2B\} & \{48, 64, 96, 128\} & \{256, 512, 1024, 1024\} & \{8, 16, 32, 32\} & \{5136, 12992, 37920, 50560\} & \{3618, 9252, 26824, 35096\}\\
		\bottomrule
	\end{tabular}
	}
	\label{tab:model-size}
\end{table*}

We solve the MILP model described in~\eqref{eqn:ilp} using the optimization solver Gurobi~\cite{gurobi}. Indicator variable $x_{ik}$ implies the placement decision of each operator. After obtaining the placement decision of operators, we employ PyTorch to implement the inter-operator model parallel inference. 

%% file: experiments.tex
\section{Experiments}
\label{sec:experiments}

In this section, we conduct extensive experiments 
to empirically evaluate \textsc{Moirai}. 
Broadly, we intend to answer the following research questions:
\begin{itemize}
\item \textbf{RQ1:} How does \textsc{Moirai} compare against the state-of-the-art approaches?
\item \textbf{RQ2:} How much does our graph coarsening method contribute to the performance of \textsc{Moirai}?
\item \textbf{RQ3:} What interesting insights and findings can we obtain from the empirical results?
\end{itemize}
Next, we present our experiment settings, 
followed by answering the above research questions one by one.
%

\subsection{Experiment Setup}
\begin{figure}[t]
	\centering
	\begin{subfigure}[t]{0.24\textwidth}
		\centering
		\includegraphics[height=4cm]{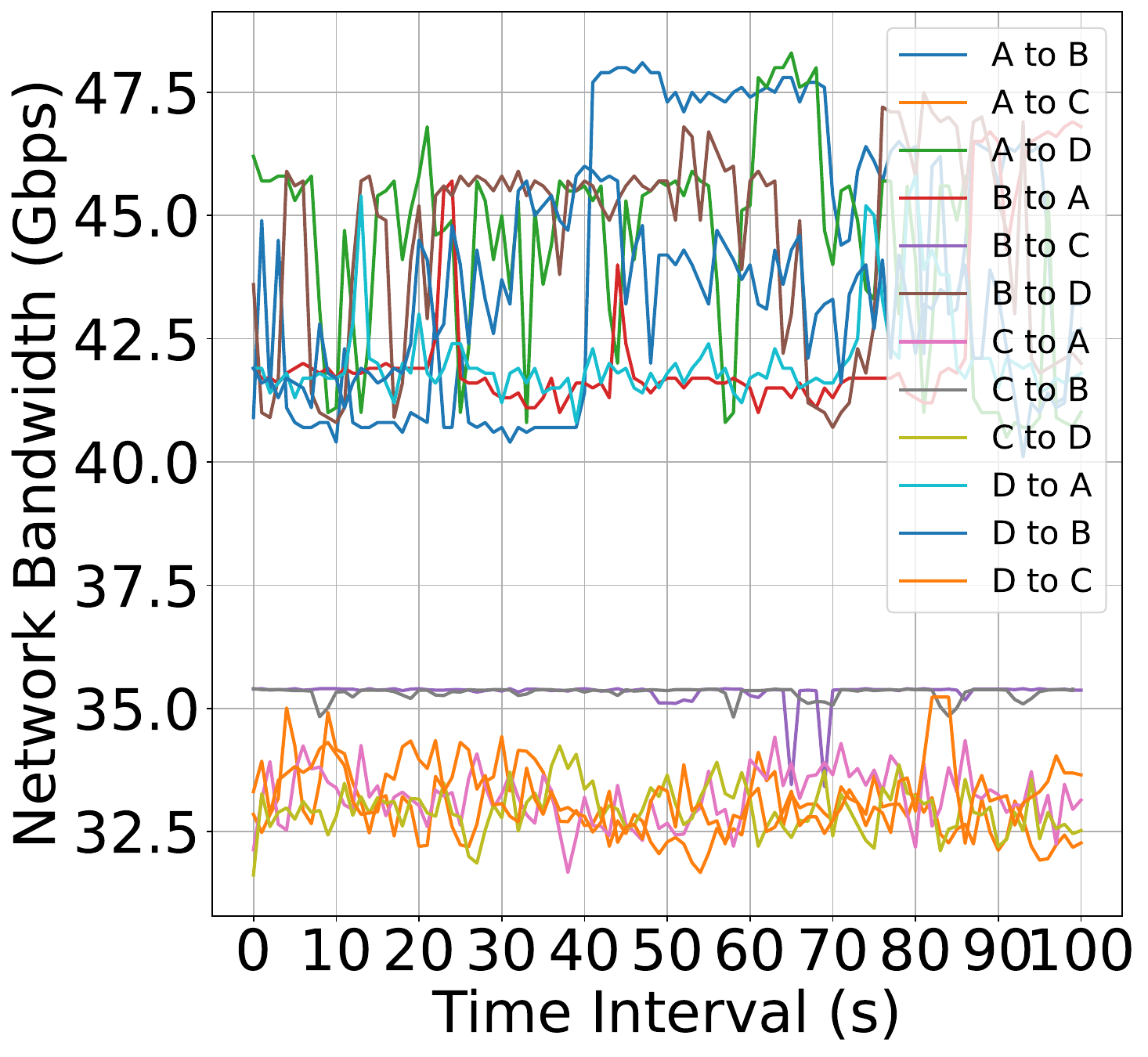}
		\caption{Inter-server scenario.}
		\label{fig:gcof-example-1}
	\end{subfigure}
	\hfill
	\begin{subfigure}[t]{0.24\textwidth}
		\centering
		\includegraphics[height=4cm]{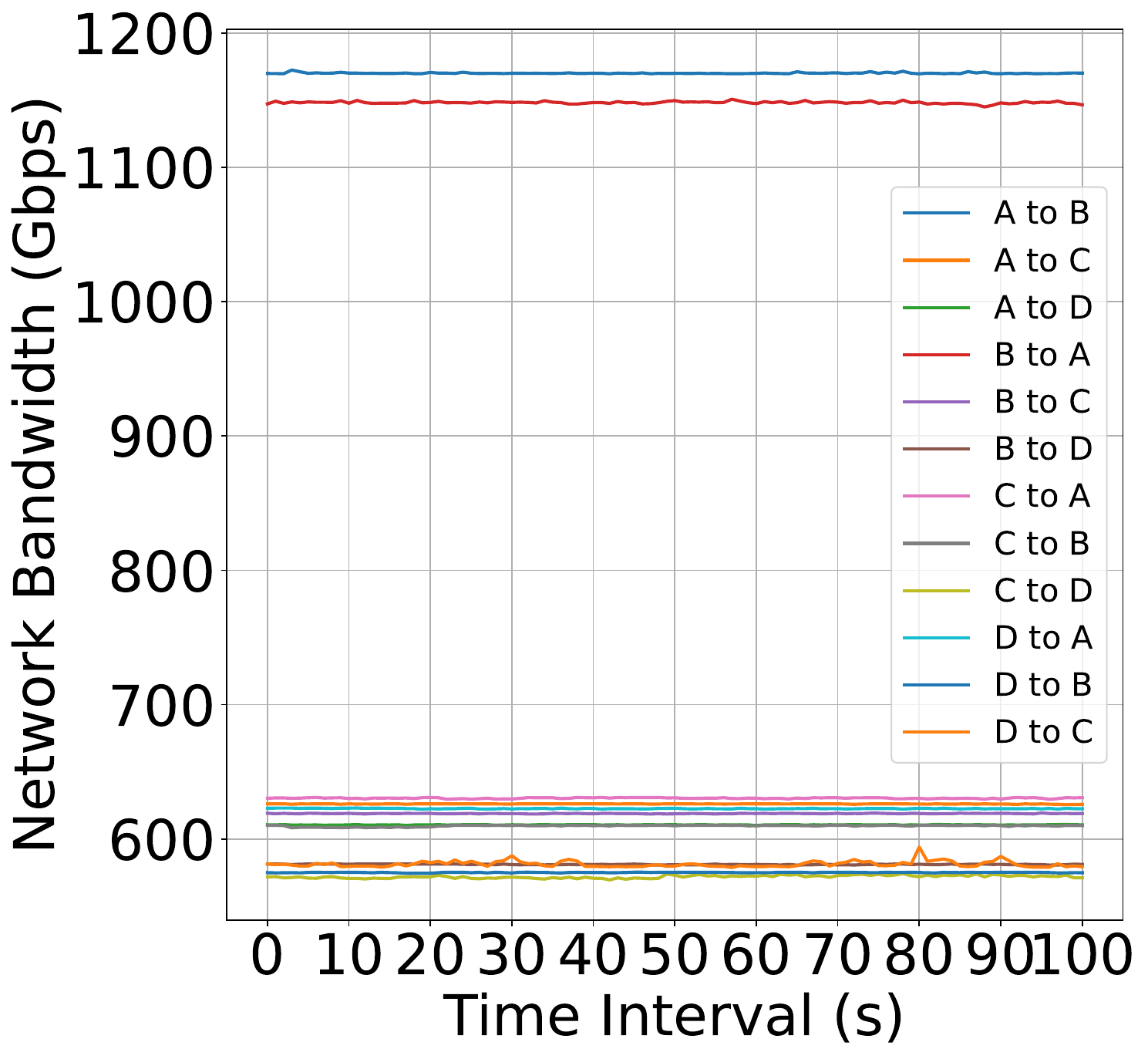}
		\caption{Intra-server scenario.}
		\label{fig:gcof-example-2}
	\end{subfigure}
	\caption{Network bandwidth between two devices over 100s.}
	\label{fig:network-bandwidth}
	\vspace{-1em}
\end{figure}

\textbf{Testbed configurations.} We demonstrate the advancement of \textsc{Moirai} through two scenarios. 
(1) \textbf{Inter-server inference:} We investigate an inter-server inter-operator model parallel inference setting, where multiple GPU servers are interconnected with a 100Gbps InfiniBand network.
(2) \textbf{Intra-server inference:} We scrutinize an intra-server inter-operator model parallel inference setting, where within each server rack, GPUs are connected via NVLink and expanded through NVSwitch to enable all-to-all communication among the GPUs.
We measured the bandwidth between every device during a 100-second period and performed calculations using the average bandwidth over this duration. 
The network bandwidth over the 100 seconds is presented in Fig.~\ref{fig:network-bandwidth}.
We list the machine configurations 
and the network conditions of the two scenarios in TABLE~\ref{tab:device-config}. 
We install NCCL~2.16 and PyTorch~v1.12 on all devices.

\textbf{Models.} We empirically evaluate \textsc{Moirai} with with three emerging models from diverse domains, including computer vision, natural language processing, and biology analysis. For each model, we apply \textsc{Moirai} to its variants in different model size that is shown in TABLE~\ref{tab:model-size}. (1) \textbf{Swin-Transformer}~\cite{liu2022swin} is a highly accurate vision model designed for image recognition. We set the resolution of input images to 1100 $\times$ 1100. (2) \textbf{GPT-3}~\cite{brown2020language} is a cutting-edge language model based on Transformer. The input of GPT-3 is word tokens. We employ a language sequence of 2048 tokens as its input. (3) \textbf{AlphaFold2}~\cite{cramer2021alphafold2} is a biological model lies in its ability to accurately predict protein structures. Following the experiment setting in~\cite{cramer2021alphafold2}, we choose the input sequence batch size of 128.

\begin{figure*}[t]
    \centering
	\begin{subfigure}[t]{0.24\textwidth}
		\centering
		\includegraphics[width=\textwidth]{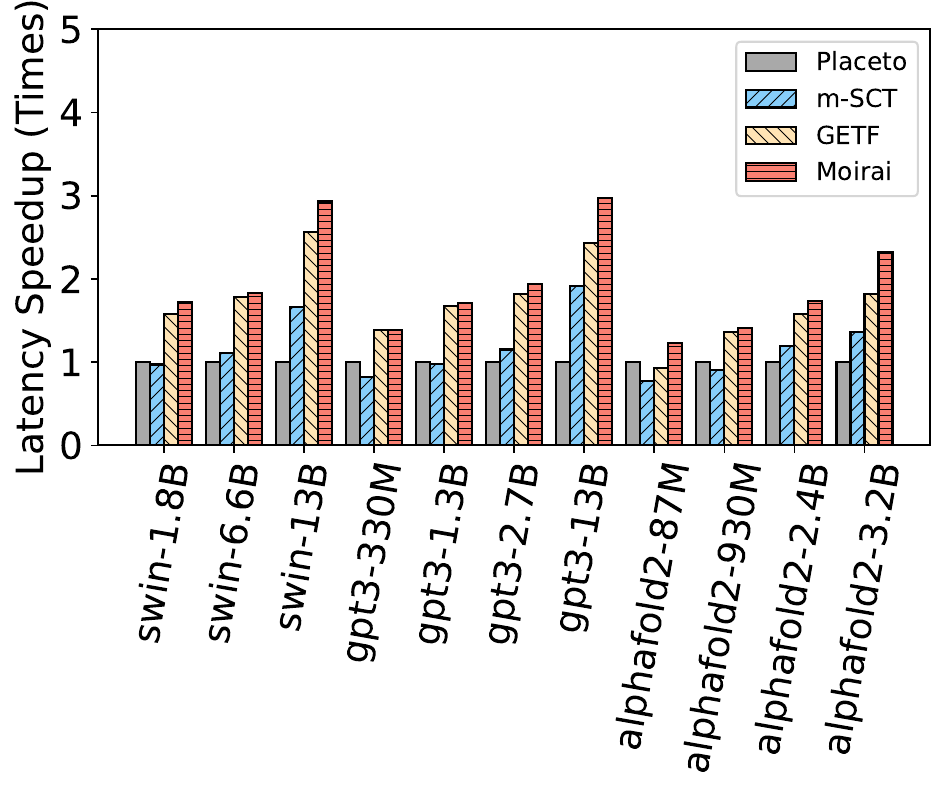}
		\caption{Inter-server scenario with original computation graphs.}
		\label{fig:inter-original}
	\end{subfigure}
	\hfill
	\begin{subfigure}[t]{0.24\textwidth}
		\centering
		\includegraphics[width=\textwidth]{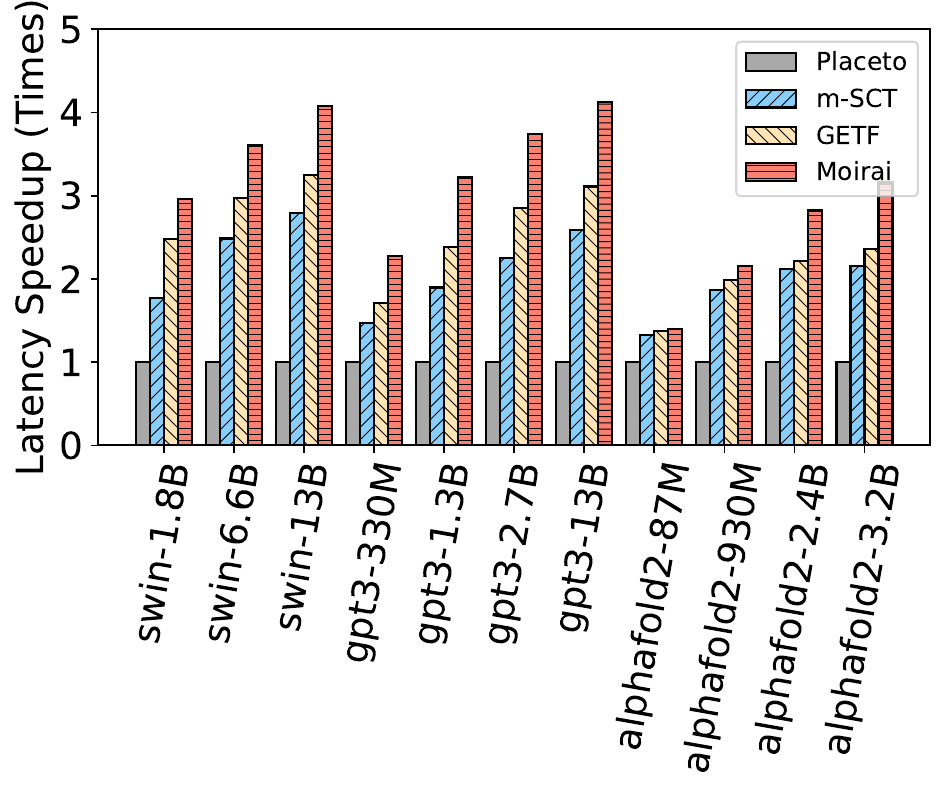}
		\caption{Intra-server scenario with original computation graphs.}
		\label{fig:intra-original}
	\end{subfigure}
	\hfill
	\begin{subfigure}[t]{0.24\textwidth}
		\centering
		\includegraphics[width=\textwidth]{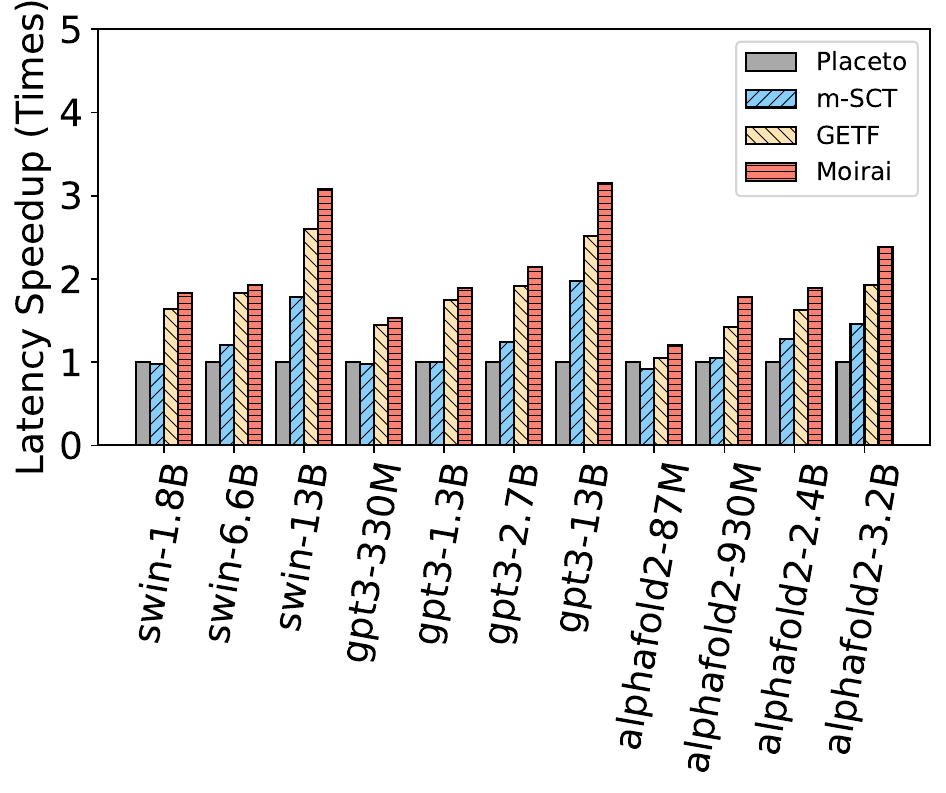}
		\caption{Inter-server scenario with coarsened computation graphs.}
		\label{fig:inter-coarsened}
	\end{subfigure}
	\hfill
	\begin{subfigure}[t]{0.24\textwidth}
		\centering
		\includegraphics[width=\textwidth]{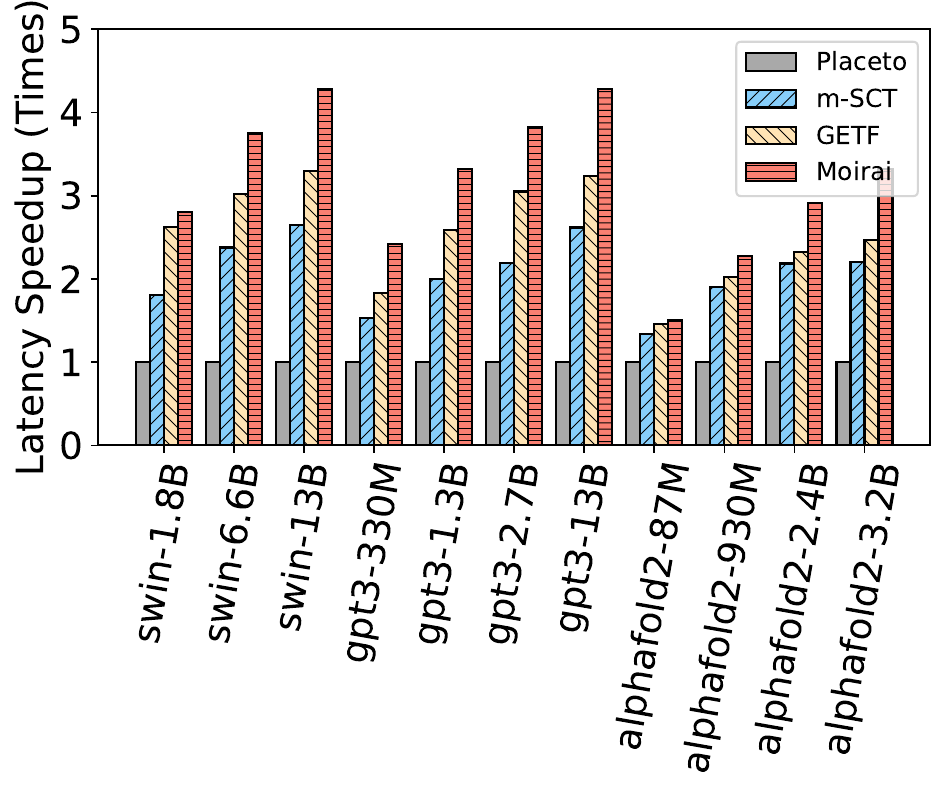}
		\caption{Intra-server scenario with coarsened computation graphs.}
		\label{fig:intra-coarsened}
	\end{subfigure}
	\caption{Inference latency speedup comparison among four algorithms.}
	\label{fig:moirai-latency-speedup}
	\vspace{-1em}
\end{figure*}

\textbf{Methods.} We compare \textsc{Moirai} with both learning-based and algorithmic methods.
(1) \textbf{Placeto}~\cite{addanki2019placeto} is a reinforcement learning approach. 
We revise its reward function to 
accommodate only the forward calculations of an input. 
Placeto serves as the baseline in our experiment.
(2) \textbf{m-SCT}, which is raised in Baechi~\cite{jeon2020baechi}, is a heuristic-based solution. 
m-SCT places operators on a device with the earliest start time 
and leverages an ILP model to find child operators. 
(3) \textbf{GETF}~\cite{su2020communication} is an exact algorithm-based method.
We implement GETF following the guidelines outlined in~\cite{su2020communication}.
We solve the GETF MILP with Gurobi.

\textbf{Metrics.} 
We employ two performance metrics to 
evaluate the inter-operator model parallel inference performance of \textsc{Moirai}.
(1) \textbf{End-to-end inference latency} 
refers to the time required for a DNN to process 
an input and generate an output. 
We deploy DNN models based on the placement 
generated by each algorithm and measure its 
end-to-end inference latency. 
To mitigate variances caused by warm-up, 
we exclude the running time of the first five batches from the measurement, 
as specified in~\cite{mirhoseini2018hierarchical}.
(2) \textbf{Placement generation time} 
denotes the duration taken by a device placement algorithm 
to generate a placement solution. We implement 
\textsc{Moirai} and its counterparts on devices equipped 
with Intel Core i7-8700 CPU and NVIDIA RTX 2080 Ti GPU, 
measuring the placement generation time. 


\begin{table}[t]
    \centering
	\setlength\extrarowheight{1.5pt}
	\caption{Placement generation time.}
	\resizebox{\columnwidth}{!}{
    \begin{tabular}{ c c c c c c c c c c }
        \toprule
        \multirow{2}{*}{\textbf{Model}} & \multirow{2}{*}{\textbf{Type}} & \multicolumn{4}{c}{\textbf{Original Computation Graph}} & \multicolumn{4}{c}{\textbf{Coarsened Computation Graph}} \\
        & & HRL & m-SCT & GETF & \textsc{Moirai} & HRL & m-SCT & GETF & \textsc{Moirai} \\
        \hline
        \multirow{3}{*}{\shortstack[c]{Swin-Transformer}} & 1.8B & 4hrs & 17.63s & 2min & 2.15min & 4hrs & 15.24s & 1.58min & 1.64min \\
        & 6.6B & 5hrs & 64.32s & 12.4min & 13.28min & 4.8hrs & 58.57s & 9.72min & 10.51min \\
        & 13B & 5hrs & 148.25s & 19.75min & 22.54min & 5hrs & 142.9s & 14.95min & 15.46min \\
        \hline
        \multirow{4}{*}{\shortstack[c]{GPT-3}} & 330M & 3.5hrs & 15.92s & 1.25min & 1.38min & 3hrs & 14.4s & 58.04s & 1.04min \\
        & 1.3B & 4hrs & 39.82s & 8.65min & 9.02min & 3.85hrs & 37.39s & 5.29min & 6.88min \\
        & 2.7B & 5hrs & 54.17s & 10.82min & 11.48min & 4.75hrs & 51.52s & 7.84min & 8.51min \\
        & 13B & 5.5hrs & 125.53s & 15.7min & 16.5min & 5hrs & 117.48s & 10.92min & 11.07min \\
        \hline
        \multirow{4}{*}{\shortstack[c]{AlphaFold2}} & 87M & 3.5hrs & 20.18s & 1.78min & 1.9min & 3.3hrs & 18.62s & 1.24min & 1.38min \\
        & 930M & 5hrs & 56.62s & 11.05min & 12.87min & 4.82hrs & 53.71s & 7.95min & 8.75min \\
        & 2.4B & 5.75hrs & 226.34s & 22.5min & 23.91min & 5hrs & 215.48s & 17.92min & 18.02min \\
        & 3.2B & 7hrs & 507.65s & 35.75min & 38.42min & 6.5hrs & 485.35s & 21.5min & 22.18min \\
        \bottomrule
    \end{tabular}
    }
    \label{tab:generation-time}
\end{table}

\subsection{Comparison with Existing Methods (RQ1)}
We compare the end-to-end inference speedup of \textsc{Moirai} 
with three counterparts under two scenarios. 
To verify the impact of the proposed graph coarsening method, 
we try applying the MILP model of \textsc{Moirai} on 
both the original DNN computation graph and the coarsened computation graph.

\textbf{End-to-end inference latency.} 
Fig.~\ref{fig:inter-original} demonstrates the end-to-end inference 
latency speedup of \textsc{Moirai} under 
inter-server scenario on original computation graphs. 
The result shows that \textsc{Moirai} reduces the end-to-end inference latency 
up to 2.98$\times$, 1.77$\times$, 1.33$\times$ 
compared to Placeto, m-SCT, and GETF respectively. 
Fig.~\ref{fig:intra-original} provides inference 
acceleration details of \textsc{Moirai} 
under intra-server scenario on original computation graphs. 
We observe that \textsc{Moirai} provides the latency speedup 
up to 4.12$\times$, 1.7$\times$, 1.35$\times$ 
compared to Placeto, m-SCT, and GETF respectively. 

Next, we are interested in 
evaluating the impact of the graph coarsening method 
of \textsc{Moirai} on reducing the end-to-end inference latency.
Fig.~\ref{fig:inter-coarsened} and~\ref{fig:intra-coarsened} 
exhibits the inference latency 
speedup of \textsc{Moirai} after 
coarsening the DNN computation graphs 
with Algorithm~\ref{alg:fusion-coarsen}. 
In the inter-server setting, 
\textsc{Moirai} outperforms Placeto, m-SCT, and GETF up to 
3.15$\times$, 1.9$\times$, and 1.25$\times$ respectively 
in reduction of the end-to-end latency.
In the intra-server setting, \textsc{Moirai} 
surpasses Placeto, m-SCT, and GETF up to 
4.28$\times$, 1.74$\times$, 1.34$\times$ respectively 
in reduction of the end-to-end latency.

\textbf{Placement generation time.}
TABLE~\ref{tab:generation-time} presents 
the placement generation time of all the approaches. 
It is observed that the HRL algorithm requires several hours 
for training and generating the final results. 
m-SCT, due to its small algorithm search space, 
requires the least amount of time. 
Considering the vast search space and limitations of the Gurobi optimizer, 
both GETF and our algorithm need minutes to generate placement. 
However, since this process is offline, 
and the optimal placement solutions of \textsc{Moirai}
are superior to those of the m-SCT, 
we believe that our approach still 
holds an advantage in the placement generation. 
Moreover, by further relaxing \textsc{Moirai} MILP model, 
we can significantly reduce the placement generation time.

\subsection{Contributions of Graph Coarsening (RQ2)}
The last two columns of TABLE~\ref{tab:model-size} show the number of operators 
in the original computation graphs 
and the number of operators in the coarsened computation graphs. 
According to the results shown in Fig.~\ref{fig:moirai-latency-speedup}, 
compared to using the original computation graph 
to generate placement results, 
the graph coarsening method has reduced 
the end-to-end inference latency by up to 5.7\% 
in the inter-server setting and up to 3.8\% 
in the intra-server setting. 
From TABLE~\ref{tab:generation-time}, 
we observe that the graph coarsening method 
has a significant effect on reducing the placement generation time, 
with an average time reduction to 71.87\% 
of the placement generation time with the original computation graphs.

\subsection{Discussion (RQ3)}
The experiments on three types of DNNs suggest 
that incorporating higher communication bandwidth 
between devices achieves better inference speedup. 
Interestingly, a point that attracts our attention is: 
we observe that as DNN models become larger, 
although the computation resource demand rises, 
large models provide greater search space, 
which increases the parallelism of the model,  
making full use of computation resources, 
and achieving a better inference acceleration with \textsc{Moirai}. 

\section{Conclusion}
\label{sec:conclusion}
In this paper, we proposed an algorithmic solution named 
\textsc{Moirai} for the device placement problem. 
\textsc{Moirai} incorporates graph optimization, 
device heterogeneity, and inter-operator model parallel inference constraints. 
We have used a graph coarsening method 
that considers runtime operator fusion 
to shrink the solution search space 
while maintaining optimality. 
The use of MILP in \textsc{Moirai} accommodates 
various constraints and can be extended to multiple heterogeneous devices. 
Extensive experiments demonstrate 
that \textsc{Moirai} consistently outperforms the state-of-the-art methods 
in reducing the end-to-end inference latency 
while ensuring a reasonable placement generation time. 
Future work involves proposing a meta-heuristic algorithm 
to expedite the placement generation process 
and providing proofs concerning 
the approximation ratio of the meta-heuristic algorithm.
